\documentclass[12pt]{iopart} 

\usepackage{graphicx}  
\usepackage{iopams}
\usepackage{caption}

\usepackage{amsopn}
\DeclareMathOperator*{\argmin}{\arg\!\min}

\begin{document}

\title[Bayesian parameter estimation of core collapse supernovae]{Bayesian parameter estimation of core collapse supernovae using gravitational wave simulations}
\author{Matthew C. Edwards$^{1,2}$, Renate Meyer$^1$ and Nelson Christensen$^2$}
\address{$^1$ Department of Statistics, University of Auckland, Auckland 1142, New Zealand}
\address{$^2$ Physics and Astronomy, Carleton College, Northfield, Minnesota 55057, USA}
\eads{\mailto{matthew.edwards@ligo.org}, \mailto{renate.meyer@auckland.ac.nz} and \mailto{nchriste@carleton.edu}}

\begin{abstract}
  Using the latest numerical simulations of rotating stellar core
  collapse, we present a Bayesian framework to extract the physical
  information encoded in noisy gravitational wave signals.  We fit
  Bayesian principal component regression models with known and
  unknown signal arrival times to reconstruct gravitational wave
  signals, and subsequently fit known astrophysical parameters on the
  posterior means of the principal component coefficients using a
  linear model.  We predict the ratio of rotational kinetic energy to
  gravitational energy of the inner core at bounce by sampling from
  the posterior predictive distribution, and find that these
  predictions are generally very close to the true parameter values,
  with 90\% credible intervals $\sim 0.04$ and $\sim 0.06$ wide for
  the known and unknown arrival time models respectively.  Two
  supervised machine learning methods are implemented to classify
  precollapse differential rotation, and we find that these methods
  discriminate rapidly rotating progenitors particularly well.  We
  also introduce a constrained optimization approach to model
  selection to find an optimal number of principal components in the
  signal reconstruction step.  Using this approach, we select 14
  principal components as the most parsimonious model.
\end{abstract}


\maketitle

\section{Introduction}

In his general theory of relativity, Albert Einstein predicted the
existence of gravitational waves (GWs) \cite{einstein:1916} ---
ripples in the fabric of spacetime caused by asymmetries in
catastrophic and highly accelerated events throughout the cosmos.
Though confirmed indirectly by observations of the binary pulsar PSR
$1913 + 16$ \cite{taylor:1989}, GWs have not been directly detected by
the global network of first generation detectors, such as Initial LIGO
in the United States \cite{abramovici:1992,ligo:2009}, Virgo in Italy
\cite{caron:1997,virgo:2012}, GEO 600 in Germany \cite{grote:2010},
and TAMA 300 in Japan \cite{ando:2005}.

The second generation of LIGO detectors, Advanced
LIGO~\cite{harry:2010}, are currently under construction and will
likely operate as early as 2015~\cite{AdvDet}.  Advanced
Virgo~\cite{AdvVirgo} should come on-line in 2016, while the Japanese
KAGRA~\cite{KAGRA} detector will join the world-wide network later in
the decade.  The ten-fold improvement in sensitivity of these
detectors \cite{harry:2010, AdvDet, smith:2009}, along with coherent
analysis between observatories, will significantly improve the chances
of detecting GWs from an astrophysical event in the Milky Way and
neighbouring galaxies.  It is therefore likely that direct detection
of GWs will occur in the near future.

Potential sources of GWs include the inspiral of compact binary star
systems (of neutron stars or black holes) followed by black hole
formation \cite{thorne:1987}, pulsars \cite{culter:2002}, rotating
core collapse supernovae (CCSN) followed by protoneutron star
formation \cite{ott:2004}, gamma-ray bursts \cite{meszaros:2006}, and
cosmic string cusps \cite{damour:2005}.

Rotating CCSN are of particular interest in this paper.  Like
neutrinos, GWs are emitted deep in the core of a progenitor and
propagate through the universe mostly unobscured by astrophysical
objects between the source and a detector on Earth. GWs act like
messengers, providing primary observables about the multi-dimensional
core collapse dynamics and emission mechanisms.  It is in this way
that GW astronomy will open a new set of eyes to view the universe,
complementing the conventional electromagnetic-type observations.

Coalescing binary star systems are the most promising source of
detectable GWs \cite{thorne:1987}, with an expected observation rate
that could be as large as a few hundred events per year for Advanced
LIGO \cite{AdvDet,abbott:2005}.  In contrast, the expected rate of
CCSN in the Milky Way is around three per century \cite{adams:2013}.
It is of great importance that appropriate data analysis techniques
are in place so we do not miss an opportunity to detect these rare
CCSN events.

The Bayesian statistical framework has proven to be a powerful tool
for parameter estimation in astrophysical and cosmological settings
\cite{loredo:1992}.  Bayesian data analysis was first introduced to
the GW community by Christensen and Meyer \cite{christensen:1998}.
Christensen and Meyer \cite{christensen:2001} then demonstrated the
usefulness of the Gibbs sampler \cite{gelman:2013, geman:1984} for
estimating five physical parameters of coalescing binary signals.
Christensen, Meyer, and Libson \cite{christensen:2004a} then went on
to show how a custom-built Metropolis-Hastings algorithm
\cite{gelman:2013, metropolis:1953, hastings:1970}, a generalization
of the Gibbs sampler, was a superior and more suited routine for
eventual implementation into the LIGO Scientific Collaboration (LSC)
algorithm library (LAL).  Parameter estimation for compact binary
inspirals has subsequently become more sophisticated in recent years
(see for example \cite{roever:2006, roever:2007a, roever:2007b,
  raymon:2009, vdsluys:2008, veitch:2010, CBC-param}).  Markov chain
Monte Carlo (MCMC) routines for inferring the physical parameters of
pulsars have also been developed \cite{christensen:2004b,
  umstatter:2004, clark:2007}.

Due to the analytical intractability and complex multi-dimensional
nature of rotating core collapse stellar events, a significant amount
of computational time must go into numerically simulating the
gravitational waveforms.  Unlike binary inspiral events, one cannot
simply use template search methods for supernova burst events as it is
computationally impossible to cover the entire signal parameter space.
It is therefore important to find alternative parameter estimation
techniques.

Summerscales \etal \cite{summerscales:2008} utilized the maximum
entropy framework to deconvolve noisy data from multiple (coherent)
detectors, with the goal of extracting a CCSN GW signal. Inference on
amplitude and phase parameters was conducted using cross correlation
between the recovered waveform and the set of competing waveforms from
the Ott \etal \cite{ott:2004} catalogue.  A match was defined as the
model with the maximum cross correlation to the recovered waveform.

Heng \cite{heng:2009} first proposed a principal component analysis
(PCA) approach to simplify the problem by reducing a given supernova
waveform catalogue space down to a small number of basis vectors.
R\"{o}ver \etal \cite{roever:2009} extended this approach and created
a novel Metropolis-within-Gibbs sampler \cite{gelman:2013} to
reconstruct test signals from the Dimmelmeier \etal catalogue
\cite{dimmelmeier:2008} in noisy data using a principal component
regression (PCR) model with random effects and unknown signal arrival
time.  They then attempted to exploit the structure of the posterior
principal component (PC) coefficients with a simple $\chi^2$ measure
of distance to determine which catalogue waveform best matched the
injected test signal.  Although the Bayesian reconstruction method
showed much promise, extraction of the underlying physical parameters
had limited success.

Logue \etal \cite{logue:2012} used nested sampling
\cite{skilling:2006} to compute Bayesian evidence for PCR models under
three competing supernova mechanisms --- neutrino, magnetorotational,
and acoustic mechanisms.  Each supernova mechanism has a noticeably
distinct gravitational waveform morphology, and the method was
successful at correctly inferring a large majority of injected
signals.  They found that for signals embedded in simulated Advanced
LIGO noise, the magnetorotational mechanism could be distinguished to
a distance of up to 10 kpc, and the neutrino and acoustic mechanisms
up to 2 kpc.

Abdikamalov \etal \cite{abdikamalov:2013} generated a new
CCSN waveform catalogue and applied matched
filtering \cite{turin:1960} to infer total angular momentum to within $\pm
20\%$ for rapidly rotating cores.  Slowly rotating cores had errors
up to $\pm 35\%$.  Along with matched filtering, they employed
the Bayesian model selection method presented in \cite{logue:2012} to
illustrate that under certain assumptions of the rotation law, the
next generation of GW detectors (Advanced LIGO, Advanced Virgo, and
KAGRA), could also extract information about the degree of precollapse
differential rotation.  The two methods worked particularly well for
rapidly rotating cores.

In this paper we present an alternative approach to parameter
estimation for rotating CCSN.  Using the Abdikamalov \etal catalogue
\cite{abdikamalov:2013}, we fit a Bayesian PCR model to reconstruct a
GW signal in noisy data.  Initially, the signal arrival time is
assumed to be known, and PC coefficients are sampled directly from the
posterior distribution.  We extend the model to incorporate an unknown
signal arrival time and construct a Metropolis-within-Gibbs MCMC
sampler (as done in \cite{roever:2009}).  We then use the posterior
means of the PC coefficients to fit the known physical parameters on
(using a linear model), and sample from the posterior predictive
distribution to make probabilistic statements about the ratio of
rotational kinetic energy to gravitational energy of the inner core at
bounce $\beta_{ic,b}$.  We apply two supervised learning algorithms
--- na\"{i}ve Bayes classifier (NBC) and $k$-nearest neighbour
($k$-NN) --- to classify the closest level of precollapse differential
rotation $A$.  We also introduce a constrained optimization approach
to model selection and attempt to find an optimal number of PCs for
the Bayesian PCR model.

The paper is organized as follows: in section 2 we describe the
simulated GW data catalogue used in our analysis;
section 3 introduces the statistical models and methods applied;
results of our analysis are presented in section 4; and a discussion
of our findings and future directions are provided in section 5.

\section{Gravitational wave data}

The waveforms used in this paper are the two-dimensional numerical
axisymmetric general-relativistic hydrodynamic rotating core collapse
and bounce supernova simulations generated by Abdikamalov \etal
\cite{abdikamalov:2013}.  Based on findings that GW signals are
essentially independent of the progenitor zero age main sequence (ZAMS)
mass by Ott \etal \cite{ott:2012}, a single presupernova progenitor
model (the 12-$M_{\odot}$ at ZAMS solar-metallicity progenitor model
from \cite{woosley:2007}) was adopted. The cylindrical rotation law
from \cite{ott:2004} was also assumed.

The GW catalogue was partitioned into a base catalogue (BC), and a
test catalogue (TC).  The BC contains $l = 92$ signals with five
levels of precollapse differential rotation $A$ (where higher values
of $A$ represent weaker differential rotation), a grid of values for
initial central angular velocity $\Omega_c$, and a grid of values for
the ratio of rotational kinetic energy to gravitational energy of the
inner core at bounce $\beta_{ic,b}$ (since $\beta_{ic,b}$ is a
function of $\Omega_c$ for a fixed progenitor structure). Each signal
in the BC was generated using the microphysical finite-temperature
Lattimer-Swesty (LS) equation of state (EOS) \cite{lattimer:1991},
parametrized deleptonization scheme from \cite{dimmelmeier:2008}, and
neutrino leakage scheme from \cite{ott:2012}.  As well as varying $A$,
$\Omega_c$, and $\beta_{ic,b}$, the TC contains 47 signals with
differing EOS and deleptonization parametrizations $Y_e(\rho)$.
Specifically, some test signals were generated using the Shen \etal
EOS \cite{shen:1998}, or an increase/decrease in $Y_e(\rho)$
parametrization by $\sim 5\%$.  The values of $\Omega_c$ and
$\beta_{ic,b}$ in the TC are in the same parameter space as those in
the BC, but with an alternative grid.  The object of our analysis is to
predict the physical parameters ($\beta_{ic,b}$ and $A$) of the
signals in the TC using information gleaned about signals in the BC.

The signals were initially sampled at 100 kHz and subsequently
downsampled by a rational factor to 16384 Hz --- the sampling rate of
the Advanced LIGO detectors.  Downsampling by a rational factor
essentially involved two steps: upsampling by an integer factor via
interpolation and then applying a low-pass filter to eliminate the
high frequency components necessary to avoid aliasing at lower
sampling rates; and downsampling by an integer factor to achieve the
desired sampling rate \cite{oppenheim:1999}.  The resampled data was
zero-bufferred to ensure each signal was the same length, $N = 16384$,
which corresponded to 1 s of data at the Advanced LIGO sampling rate.
Each signal was then aligned so that the first negative peak (not
necessarily the global minimum), corresponding to the time of core
bounce, occurred halfway through the time series.

In this analysis, the source of a GW emission is assumed to be
optimally oriented (perpendicular) to a single interferometer.  Each
signal is linearly polarized with zero cross-polarization.

\begin{center}
  \includegraphics[width=1\linewidth]{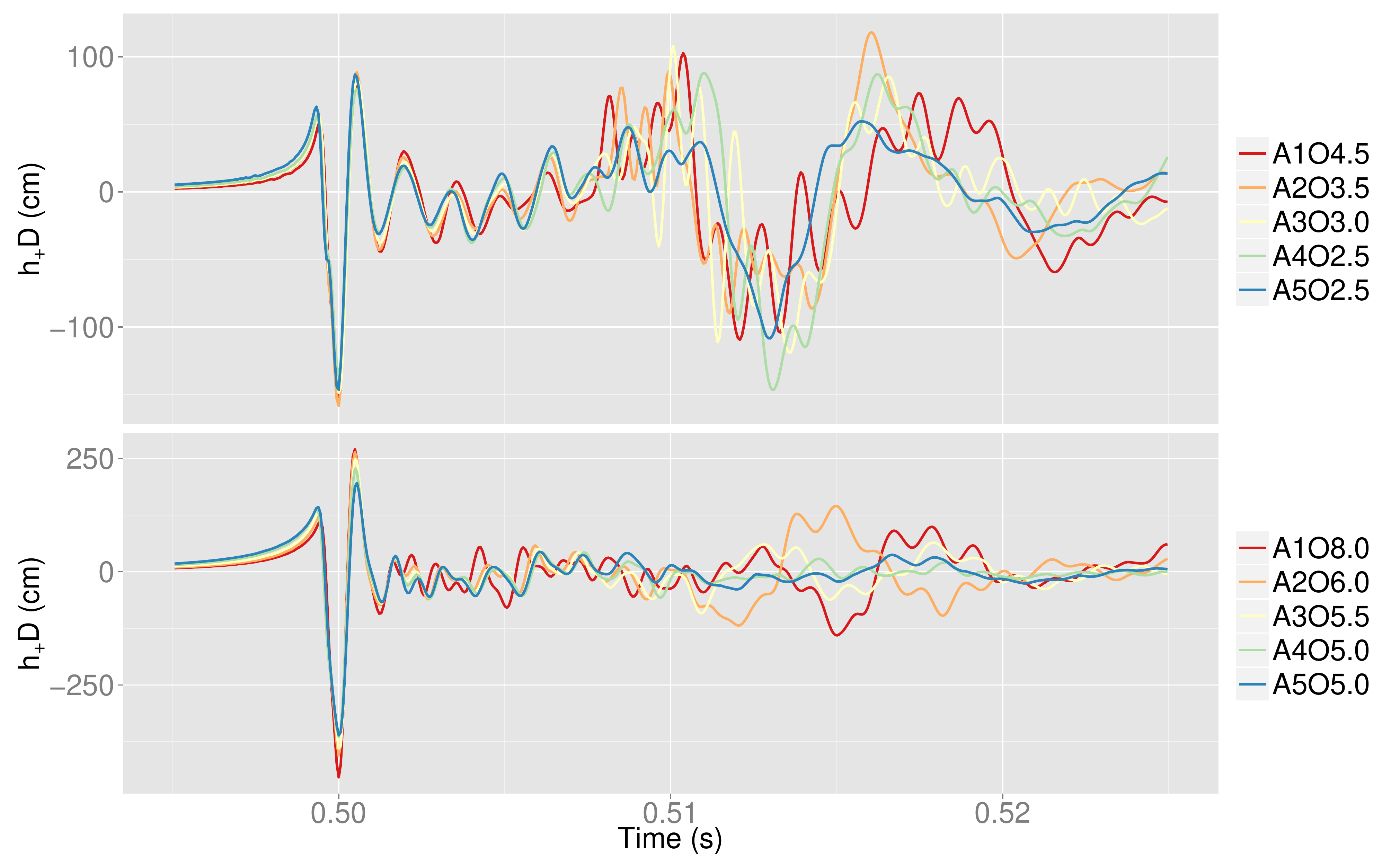}
  \captionof{figure}{A snapshot of the Abdikamalov \etal
    \cite{abdikamalov:2013} catalogue. The top panel shows the GW
    strain (scaled by source distance) for five models with different
    levels of precollapse differential rotation (from strongest
    differential rotation $A1$ to weakest $A5$), each with
    $\beta_{ic,b} \sim 0.03$ (i.e., slowly rotating progenitors).  The
    bottom panel is the same, but for rapidly rotating progenitors
    with $\beta_{ic,b} \sim 0.09$.}
  \label{fig:gw}
\end{center}

We can see a general waveform morphology in figure~\ref{fig:gw}.
During core collapse, there is a slow increase in GW strain until the
first local maximum is reached (before 0.5 s).  This is followed by
core bounce, where the strain rapidly decreases towards a local
minimum (at 0.5 s).  This corresponds to the time when the inner core
expands at bounce.  After this, there is a period of ring-down
oscillations.  For slowly rotating progenitors, we see in the top
panel of figure~\ref{fig:gw} that the GW strain is essentially the
same during collapse and bounce and only differs during ring-down.  For
the rapidly rotating progenitors presented in the bottom panel of
figure~\ref{fig:gw}, larger precollapse differential rotation results
in: a smaller local maximum during core collapse; a more negative
local minimum during core bounce; and a larger first ring-down peak.
Because of these patterns, Abdikamalov \etal \cite{abdikamalov:2013}
concluded that inferences about precollapse differential rotation
could in principal be made for rapidly rotating cores.

The data analyzed are CCSN GW signals injected in coloured Gaussian
noise using the Advanced LIGO noise curve with one-sided power
spectral density (PSD), $S_1(f)$.  The data is then Tukey windowed to
mitigate spectral leakage.  Rather than fixing source distance
to 10 kpc (as done in \cite{abdikamalov:2013}), this analysis assumes
a fixed signal-to-noise ratio (SNR) of $\rho = 20$.  SNR is defined as
\begin{equation}
  \label{eq:SNR}
  \rho = \sqrt{4 \sum_j \frac{\frac{\Delta_t}{N}|\tilde{y_j}|^2}{S_1(f_j)}} ,
\end{equation}
where $\tilde{y_j}, j = 1, 2, \ldots, N$, are the Fourier transformed
data, $\Delta_t$ is the distance between two consecutive time points,
and $f_j, j = 1, 2, \ldots, N$, are the Fourier frequencies.  As done
in \cite{roever:2009}, $S_1(f)$ is estimated \textit{a priori} by
averaging 1000 empirical periodograms from identically simulated
Advanced LIGO noise.  This corresponds to a realistic scenario where
the noise spectrum must be estimated as well.

Although supernovae from the Milky Way will not produce SNRs as small
as $\rho = 20$, we choose this value to illustrate that our methods
are robust at lower SNRs.

\section{Methods and models}

\subsection{Bayesian inference}

  Bayesian inference requires three pivotal quantities.
  The \textit{likelihood} function $p(\mathbf{z} | \boldsymbol\theta)$
  is the probability density function (PDF) of the data $\mathbf{z}$,
  conditional on the random vector of model parameters
  $\boldsymbol\theta$.  The \textit{prior} $p(\boldsymbol\theta)$ is
  the PDF of the model parameters, that takes into account all of the
  information known about $\boldsymbol\theta$ before the data is
  observed.  The \textit{posterior} $p(\boldsymbol\theta |
  \mathbf{z})$ is the updated PDF of model parameters after the data is
  observed.  These quantities are related via Bayes' theorem
\begin{eqnarray}
  \label{eq:bayes}
  p(\boldsymbol\theta | \mathbf{z}) 
& = & \frac{p(\boldsymbol\theta)p(\mathbf{z}|\boldsymbol\theta)}{m(\mathbf{z})} \\
& \propto & p(\boldsymbol\theta)p(\mathbf{z} | \boldsymbol\theta) ,
\end{eqnarray}
where $m(\mathbf{z}) = \int p(\boldsymbol\theta)p(\mathbf{z} |
\boldsymbol\theta) \mathrm{d}\boldsymbol\theta$ is the
\textit{marginal likelihood} and is treated as a normalizing constant
since it is independent of $\boldsymbol\theta$.  That is, the
posterior is proportional to the prior multiplied by the likelihood.

Posterior sampling can be performed directly if the posterior PDF has
a closed analytical form.  Otherwise, MCMC
techniques are a useful work-around.  The key building blocks in MCMC
simulations are the Gibbs sampler \cite{geman:1984} and the
Metropolis-Hastings algorithm \cite{metropolis:1953, hastings:1970}.
We use a combination of the two --- the so-called
Metropolis-within-Gibbs sampler --- in this study.  For a detailed
account of Bayesian inference and MCMC algorithms, refer to
\cite{gelman:2013}.


\subsection{Model 1: Bayesian PCR with known signal arrival time}


We aim to first reduce the dimension of the BC by a PCA, or
equivalently a singular value decomposition (SVD) as suggested by Heng
\cite{heng:2009}.  Each BC waveform is represented as a linear
combination of orthonormal basis vectors, where the projection of the
data onto the first basis vector has maximum variance, the projection
onto the second basis vector has second highest variance, and so on.
By considering only projections on the first $d < l$ basis vectors,
the so-called $d$ PCs, a parsimonious representation of the catalogue
signals in $d$ dimensions is achieved that preserves as much of the
information of the original BC as possible.

Once PCA is conducted, the first $d$ PCs are treated as the
explanatory variables of a linear model.  The data analyzed is a time
series vector $\mathbf{y}$ of length $N$ and decomposes into additive
signal and noise components.  Let $\tilde{\mathbf{y}}$ be the Fourier
transformed data vector of length $N$ and let $\tilde{\mathbf{X}}$ be
the $N \times d$ design matrix, whose columns are the Fourier
transformed mean-centered PC vectors from the BC.  The frequency
domain linear model is
  \begin{equation}
    \label{eq:freqLM}
    \tilde{\mathbf{y}} = \tilde{\mathbf{X}} \boldsymbol\alpha + \tilde{\boldsymbol\epsilon},
  \end{equation}
  where $\boldsymbol\alpha$ is the vector of PCR
  coefficients and $\tilde{\boldsymbol\epsilon}$ is the Fourier
  transformed coloured zero-mean Gaussian noise vector whose variance
  terms are proportional to the \textit{a priori known} one-sided
  power spectral density $S_1(f)$.  That is,
  \begin{equation}
    \label{eq:variances}
    \sigma^2_{f_j} = \frac{N}{4 \Delta_t} S_1(f_j).
  \end{equation}
  Due to Hermitian symmetry, the frequency domain data vector
  $\tilde{\mathbf{y}}$ contains only the non-redundant real and
  imaginary components and is therefore the same length as the time
  domain vector $\mathbf{y}$.  Conversion between time and frequency
  domains is conducted using a fast Fourier transform (FFT).

  The likelihood for the Bayesian PCR model with known signal arrival
  time is
\begin{equation}
  \label{eq:likelihood1}
  p(\tilde{\mathbf{y}} | \boldsymbol\alpha) \propto \exp \left( -2 \sum_{j = 1}^{N} \frac{\frac{\Delta_t}{N}\left( \tilde{y}_j - \left( \tilde{\mathbf{X}} \boldsymbol\alpha \right)_j \right)^2}{S_1 \left(f_j\right)} \right).
\end{equation}
Assuming flat ($\mathrm{Uniform}(-\infty, \infty)$) priors on
$\boldsymbol\alpha$, the posterior distribution for the PC
coefficients is
\begin{equation}
  \label{eq:condpost1}
  \mathrm{P}(\boldsymbol\alpha | \tilde{\mathbf{y}}) = \mathrm{N}(\boldsymbol\mu, \boldsymbol\Sigma),
\end{equation}
where
\begin{eqnarray}
  \boldsymbol\Sigma &=& (\tilde{\mathbf{X}}^{'} \mathbf{D}^{-1} \tilde{\mathbf{X}})^{-1}, \\
  \boldsymbol\mu &=& \boldsymbol\Sigma \tilde{\mathbf{X}}^{'} \mathbf{D}^{-1} \tilde{\mathbf{y}},
\end{eqnarray}
and $\mathbf{D} = \mathrm{diag}(\sigma^2_{f_j})$ is the diagonal
covariance matrix of individual variances for the noise component.
This multivariate normal distribution can be sampled directly with no
MCMC required.

Noninformative priors were chosen for this model.  It was important to
keep the data and prior knowledge separate and distinct, and to avoid
using information from the waveform catalogue for both purposes.  As
the only data available for analysis were the generated GWs, we
assumed complete prior ignorance on all model parameters.

\subsection{Model 2: Bayesian PCR with unknown signal arrival time}

The Bayesian PCR model presented in the previous section assumed a
known signal arrival time.  The precise arrival time of a GW signal to
an interferometer will generally not be known in practice,
and must therefore be included as an additional unknown parameter in
the statistical model. 

Let $T$ be a cyclical time shift representing
the unknown signal arrival time, and let $\tilde{\mathbf{X}}_T$ be the
Fourier transformed design matrix $\tilde{\mathbf{X}}$ shifted by lag
$T$, such that the Fourier transformed PCs are aligned with the
Fourier transformed data vector, $\tilde{\mathbf{y}}$.  This
transformation can be done directly in the frequency domain as a phase
shift by multiplying the columns of $\tilde{\mathbf{X}}$ by $\exp(-2
\pi \mathrm{i} f T)$.

We build on the Bayesian signal reconstruction model presented in
\cite{roever:2009}, although our primary goal is inferring the
physical parameters of a supernova progenitor and not signal
reconstruction.  

Using the same reasoning described in the previous section, we assume
flat priors on $\boldsymbol\alpha$ and $T$.  The likelihood for the
Bayesian PCR model with unknown signal arrival time is
\begin{equation}
  \label{eq:likelihood2}
  p(\tilde{\mathbf{y}} | \boldsymbol\alpha, T) \propto \exp \left( -2 \sum_{j = 1}^{N} \frac{\frac{\Delta_t}{N}\left( \tilde{y}_j - \left( \tilde{\mathbf{X}}_T \boldsymbol\alpha \right)_j \right)^2}{S_1 \left(f_j\right)} \right).
\end{equation}
For a given time shift $T$, the conditional posterior
distribution for the PC coefficients $\boldsymbol\alpha | T$ is
\begin{equation}
  \label{eq:condpost2}
  \mathrm{P}(\boldsymbol\alpha | T, \tilde{\mathbf{y}}) = \mathrm{N}(\boldsymbol\mu_T, \boldsymbol\Sigma_T),
\end{equation}
where
\begin{eqnarray}
  \boldsymbol\Sigma_T &=& (\tilde{\mathbf{X}}^{'}_T \mathbf{D}^{-1} \tilde{\mathbf{X}}_T)^{-1}, \\
  \boldsymbol\mu_T &=& \boldsymbol\Sigma_T \tilde{\mathbf{X}}^{'}_T \mathbf{D}^{-1} \tilde{\mathbf{y}} .
\end{eqnarray}
To estimate $\boldsymbol\alpha$ and $T$, we construct a Markov chain
whose stationary distribution is the posterior distribution of
interest using Metropolis-within-Gibbs sampler \cite{gelman:2013}.
This is essentially a Gibbs sampler that alternates between the full
set of conditional posterior distributions $P(\boldsymbol\alpha | T,
\tilde{\mathbf{y}})$ and $P(T | \boldsymbol\alpha,
\tilde{\mathbf{y}})$.  The former can be sampled directly using
equation~(\ref{eq:condpost2}), and the latter requires a random walk
Metropolis step, hence the name Metropolis-within-Gibbs.

After initialization, step $i + 1$ in the Metropolis-within-Gibbs algorithm is:
\begin{enumerate}
\item Directly sample the conditional posterior of $\alpha_{i+1} |
  T_i$ using equation~(\ref{eq:condpost2});
\item Propose $T_{*}$ from $t_{\nu}(T_i, \zeta^2)$ and accept $T_{i+1}
  = T_{*}$ with the Metropolis acceptance probability
  \begin{equation}
    \label{eq:metropolis}
    r = \min\left(1, \frac{p(T_{*} | \boldsymbol\alpha, \tilde{\mathbf{y}})}{p(T_i | \boldsymbol\alpha, \tilde{\mathbf{y}})}\right).
  \end{equation}
  Otherwise reject and set $T_{i+1} = T_i$.
\end{enumerate}

A $t$-distribution was chosen as the proposal distribution for the
algorithm.  It has a similar (symmetrical) shape to the normal
distribution but has heavier tails and an additional
degrees-of-freedom parameter, $\nu$.  The heavier tails of the
$t$-distribution results in bolder and more robust proposals than the
normal distribution, ensuring the algorithm does not get stuck in
local modes \cite{gelman:2013}.  The degrees-of-freedom parameter was
set to $\nu = 3$, which is the smallest integer that yields a
distribution with finite variance.  The proposal for $T_{i+1}$ is
centered on $T_i$, and has scale parameter $\zeta^2$ that is initially
and arbitrarily set to 0.05, and subsequently automatically tuned
during the algorithm to ensure good mixing and acceptance rates.


\subsection{Posterior predictive distribution}

For each of the $l = 92$ signals in the BC and $m = 47$ signals in the
TC, we fit both Bayesian PCR models, with $d$ PCs (where the choice of
$d$ is explained below).  We then construct an $l \times (d + 1)$
design matrix $\mathbf{A}$ whose rows are the \textit{posterior means}
of the $d$ PC coefficients, plus an intercept term, for each of the
$l$ signals in the BC.  The primary goal is to exploit the posterior
PC coefficient space to make inferences on the physical parameters of
rotating core collapse stellar events in the TC.  We accomplish this
by fitting a linear model with the known physical parameters from the
BC as the response variable on the design matrix $\mathbf{A}$ using
  \begin{equation}
    \label{eq:LM2}
    \boldsymbol\xi = \mathbf{A} \boldsymbol\gamma + \boldsymbol\delta,
  \end{equation}
  where $\boldsymbol\xi$ is the vector of known continuous physical
  parameters, $\boldsymbol\gamma$ is the vector of regression
  coefficients, and $\boldsymbol\delta$ is an error term.  The error
  term is assumed to come from an independent and identically
  distributed normal distribution with zero mean and variance
  $\sigma^2$.  Predictions using the \textit{posterior predictive
    distribution} are the primary interest in this analysis, and not the
  model parameters themselves.

  Assuming the convenient noninformative prior distribution that is
  uniform on $(\boldsymbol\gamma, \log \sigma)$, the posterior
  predictive distribution for a normal linear model is a multivariate
  $t$-distribution and can be sampled from directly with no MCMC
  \cite{gelman:2013}.  The formula is
  \begin{equation}
    \mathrm{P}(\check{\boldsymbol\xi} | \boldsymbol\xi) = t_{l - d}\left(\check{\mathbf{A}}\hat{\boldsymbol\gamma}, s^2\left(I + \check{\mathbf{A}} \mathbf{V}_{\boldsymbol\xi} \check{\mathbf{A}}^{'}\right)\right) ,
  \end{equation}
  where $\check{\boldsymbol\xi}$ is the vector of outcomes we wish to
  predict (i.e., the physical parameters from signals in the TC),
  $\check{\mathbf{A}}$ is the $m \times (d + 1)$ matrix whose rows are
  the posterior means of the signals in the TC (and an intercept term)
  from the Bayesian PCR step, $I$ is the $m \times m$ identity matrix,
  and
\begin{eqnarray}
  \mathbf{V}_{\boldsymbol\xi} &=& (\mathbf{A}^{'}\mathbf{A})^{-1}, \\
  \hat{\boldsymbol\gamma} &=& \mathbf{V}_{\boldsymbol\xi} \mathbf{A}^{'}\boldsymbol\xi, \\
  s^2 &=& \frac{1}{l - d} (\boldsymbol\xi - \mathbf{A}\hat{\boldsymbol\gamma})^{'}(\boldsymbol\xi - \mathbf{A}\hat{\boldsymbol\gamma}).
\end{eqnarray}

\subsection{Deviance information criterion and constrained optimization}

The \textit{deviance} is defined as $D = -2 \log p(\mathbf{z} |
\boldsymbol\theta)$ where $p(\mathbf{z} | \boldsymbol\theta)$ is the
likelihood of a statistical model, and $\boldsymbol\theta$ is the
vector of model parameters.  The \textit{deviance information
  criterion} (DIC) is a Bayesian model comparison technique and a
generalization of Akaike information criterion (AIC) for hierarchical
models \cite{spiegelhalter:2002}.  DIC is defined as
\begin{eqnarray}
  \mathrm{DIC} & = & \bar{D} + p_D  \label{eqn:DIC1} \\
  & = & 2\bar{D} - D(\bar{\boldsymbol\theta}) , \label{eqn:DIC2}
\end{eqnarray}
where $\bar{D}$ is the mean deviance from posterior samples, $p_D$ is
the effective number of parameters, and $D(\bar{\boldsymbol\theta})$
is the deviance evaluated at the posterior means of the
parameters. When comparing competing statistical models, the lowest
DIC is preferred. $\bar{D}$ is a measure of fit, and $p_D$ is a
measure of model complexity used to penalize models with too many
parameters.  Equation~(\ref{eqn:DIC1}) therefore illustrates how DIC
incorporates Occam's razor, allowing one to select a parsimonious
model, balancing between fit and complexity.  Equation~(\ref{eqn:DIC2}),
on the other hand, provides a simple method for computing DIC.
$\bar{D}$ is calculated by evaluating the deviance for each of the
stored model parameters $\boldsymbol\theta$ that have been sampled
from their joint posterior PDF, and then taking the average.
$D(\bar{\boldsymbol\theta})$ is calculated by finding the posterior
mean of each of the model parameters $\bar{\boldsymbol\theta}$ and
then evaluating the deviance.

DIC is the preferred model comparison technique in this analysis. A
popular alternative, Bayes factors, would require computing the
marginal likelihood from equation~(\ref{eq:bayes}), which involves
multi-dimensional integration over a large number of parameters.
Numerical techniques such as nested sampling \cite{skilling:2006} can
be used to derive the marginal likelihood but these methods require
significant computational time and power.  On the other hand, DIC is
easily computed from posterior samples.  Another benefit of using DIC
over Bayes factors is that improper priors (which we have assumed in
this analysis) do not violate any conditions of use.  Bayes factors,
on the other hand, are no longer applicable when improper priors are
used.

The choice of the number of PCs has been arbitrary in most of the
supernova GW parameter estimation literature and this number has
usually been $d = 10$ (see for example \cite{roever:2009,
  abdikamalov:2013}).  We propose a method for selecting the optimal
choice of $d$ based on careful analysis of the DIC for competing
models and constrained optimization. Since PCs are ordered by the
total amount of variation they make up in the data set, PCA provides a
convenient ordering system for nested modelling.  Let $M_d, d \in \{1,
2, \ldots, 92\}$, represent the set of possible PCR models,
where $d$ is the number of explanatory variables.  The models are
nested such that $M_1$ has one explanatory variable (PC1), $M_2$ has
two explanatory variables (PC1 and PC2), and so on.

For each of the $l = 92$ signals in the BC (injected in Advanced LIGO
noise), all of the models $M_d, d \in \{1, 2, \ldots, 92\}$, are
fitted and then compared using DIC.  The model with the lowest DIC is
the best fit to the data.  However, models with an absolute
difference in DIC of $\lesssim 5$ are generally taken to be
indistinguishable from one another \cite{spiegelhalter:2002} and so to
prevent over-fitting, we propose a constrained optimization routine,
where we select the smallest $d$ such that the difference in DIC
between $M_d$ and the model with the minimum DIC is less than 5.  More
specifically, let $M_{\min}$ be the model with the minimum DIC, then
find $d$ such that
\begin{equation}
  \label{eq:optimization}
  \argmin_d \bigg\{ \mathrm{DIC}(M_d) - \mathrm{DIC}(M_{\min}) < 5 \bigg\} .
\end{equation}

We employ this routine for each of the $l = 92$ BC signals, and look
at the distribution of $M_d$'s over all signals.  The median of this
distribution seems a prudent choice for a general-purpose number of
PCs since these distributions tend to be skewed. It is
important to note here that we cannot choose a different value for $d$
for each signal when implementing these models as this would lead to a
very sparse design matrix $\mathbf{A}$ when sampling from the
posterior predictive distribution.

We refer the reader to figure~\ref{fig:occam} in the results section
of this paper for an example of this method in action.

\subsection{Na\"{i}ve Bayes classifier}

The NBC \cite{ripley:1996} is a common supervised learning
algorithm and discriminant method used to group objects into a
discrete set of classes based on a set of features.  The algorithm
requires a \textit{training set} of objects with known groupings and
observed features.  Once the algorithm has learnt from the training
set, each object in a \textit{test set} (containing a set of observed
features and potentially unknown classes) is assigned to the group
that it has the highest probability of belonging to.

The ``Bayes'' component of the method refers to Bayes' theorem
\begin{equation}
  \label{eq:naive}
  p(c|\mathbf{u}) \propto p(c) p(\mathbf{u} | c)
\end{equation}
where $c \in C$ is the class that an object could belong to, and
$\mathbf{u}$ are the features we wish to exploit to classify the 
object.  That is, given some observed features $\mathbf{u}$, what is
the posterior probability of an object belonging to class $c$?

The ``na\"{i}ve'' component refers to the assumption of conditional
independence of the model features $\mathbf{u} = (u_1, u_2, \ldots,
u_d)$.  This assumption implies the joint PDF $p(\mathbf{u} | c)$ can
be factorized as the product of marginal distributions
\begin{equation}
  \label{eq:factorize}
  p(\mathbf{u} | c) = \prod_{i = 1}^d p(u_i | c),
\end{equation}
and so equation~(\ref{eq:naive}) becomes
\begin{equation}
  \label{eq:naive2}
  p(c | \mathbf{u}) = p(c) \prod_{i = 1}^d p(u_i | c).
\end{equation}
Given class $c$, each feature $(u_1, u_2, \ldots, u_d)$ is assumed to
be independently normally distributed.  The model parameters are
approximated using the relative frequencies from the training set.
The class prior probabilities $p(c)$ are specified as the number of
objects in class $c$ in the training set divided by the total number
of objects.  Objects are grouped into the class that yields the
highest posterior probability.  This is known as the maximum \textit{a
  posteriori} (MAP) decision rule.

\subsection{$k$-nearest neighbour}

An alternative machine learning algorithm to the NBC is the $k$-NN
\cite{ripley:1996}, which uses a measure of ``closeness'' between
objects rather than a probabilistic framework.  We choose $k = 1$,
meaning that an object in the test set is assigned to the class of its
single nearest neighbour in the training set.  Ties in distance are
settled at random.

The definition of closeness in this context depends on the choice
of metric.  As commonly used in the literature \cite{ripley:1996}, a
Euclidean distance is assumed.  For any object with features
$\mathbf{u} = (u_1, u_2, \ldots, u_d)$ in the test set, the $k$-NN
algorithm finds the object with features $\mathbf{v} = (v_1, v_2,
\ldots, v_d)$ in the training set that minimizes the Euclidean
distance
\begin{equation}
  \label{eq:euclidean}
  \mathrm{distance}(\mathbf{u}, \mathbf{v}) = \sqrt{\sum_{i = 1}^d (u_i - v_i)^2},
\end{equation}
and then assigns $\mathbf{u}$ to the class of $\mathbf{v}$.

\section{Results}

\subsection{Model selection}

An important statistical task is to select a prudent number of model
dimensions whilst incorporating Occam's razor into the decision making
process.  More specifically, one needs to balance model fit against
complexity to ensure there is no over-fitting.  In the context of PCA,
the decision is usually made based on the amount of variation the
first $d$ PCs contribute to the data set (i.e., analyzing Scree
plots).  This approach is arbitrary and deals specifically with
dimension reduction, but not Occam's razor.  We propose an alternative
approach, involving DIC and constrained optimization.

We analyze the change in DIC as model dimensionality
increases. Figure~\ref{fig:occam} illustrates DIC as a function of
model dimensionality for signal $A1O2.5$ from the Abdikamalov \etal
catalogue \cite{abdikamalov:2013}.  This is the typical shape of the
DIC curve for all signals in the BC and a good visual aid of Occam's
razor in action.  There tends to be a sharp decrease in DIC as the
model dimension increases at the beginning, where model fit is
improving.  DIC flattens out and then reaches a minimum, where there
is the best balance between fit against complexity.  After this, there
is a slow rise in DIC as the model dimension increases and becomes too
complex.

\begin{center}
  \includegraphics[width=0.9\linewidth]{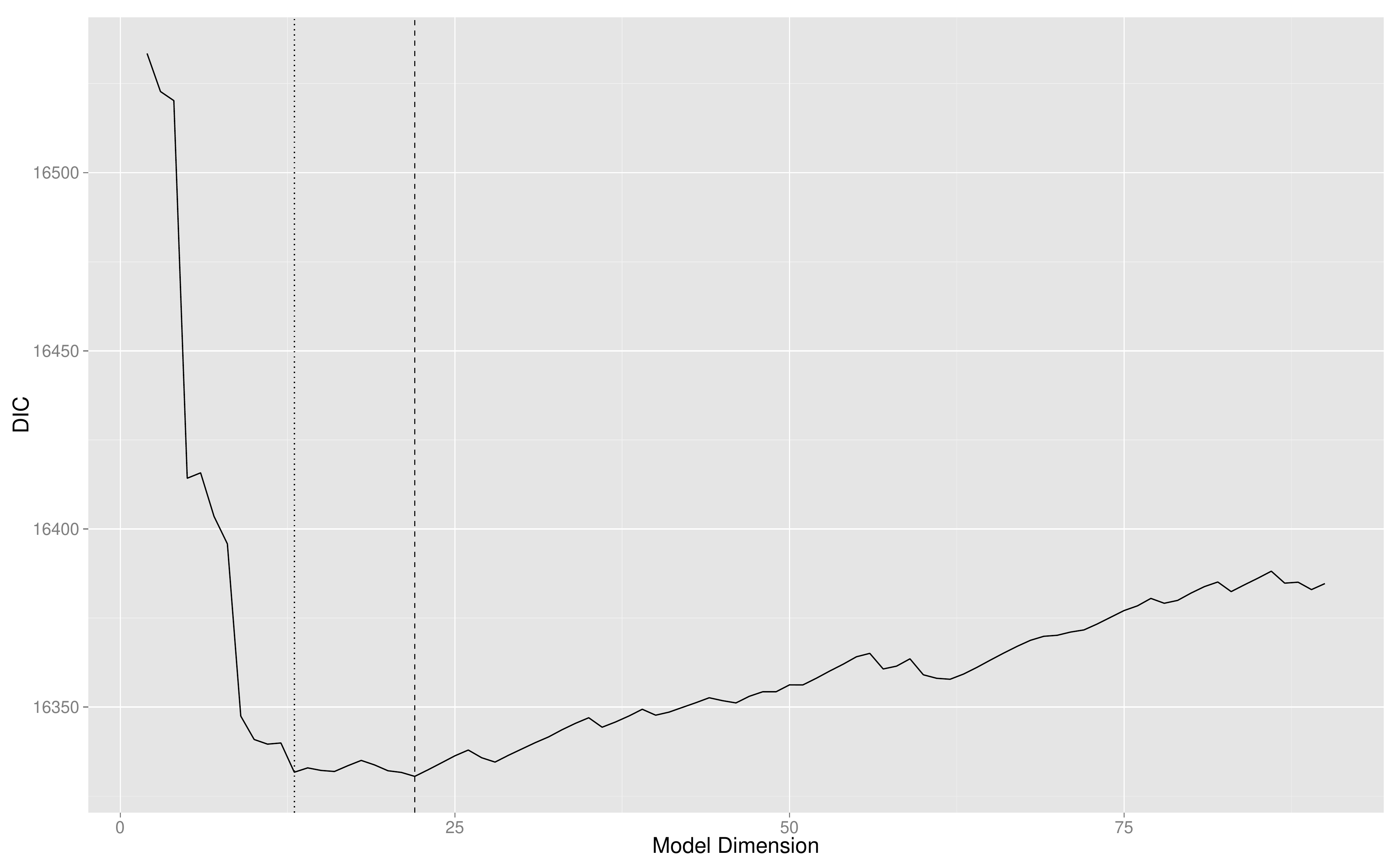}
  \captionof{figure}{DIC as a function of model dimensionality for
    model $A1O2.5$ from the Abdikamalov \etal catalogue
    \cite{abdikamalov:2013}.  The dashed vertical line to the right
    represents the model with the minimum DIC ($M_{\min} = M_{22}$).
    The dotted vertical line to the left represents the model
    dimension after constrained optimization ($M_d = M_{13}$).}
  \label{fig:occam}
\end{center}

The flat basin around the global minimum in figure~\ref{fig:occam} is
of particular interest.  Since models with an absolute difference in
DIC of less than 5 are essentially indistinguishable, it is sensible
to select the model with the smallest number of dimensions in this
region to prevent over-fitting.  For signal $A1O2.5$, we see a
significant decrease in model dimensionality from $M_{\min} = M_{22}$
to $M_d = M_{13}$.  The choice of $d$ for this particular signal is $d
= 13$.

It is important to note that $d$ will differ between GW signals but we
must only choose one general-purpose value of this.  We therefore
conduct the proposed constrained optimization model selection method
on all of the $l = 92$ BC signals and take the median of the
distribution of $d$'s as the general-purpose $d$.  We prefer the
median to the mean as our central measure as it is more robust against
outliers.

\begin{center}
  \includegraphics[width=1\linewidth]{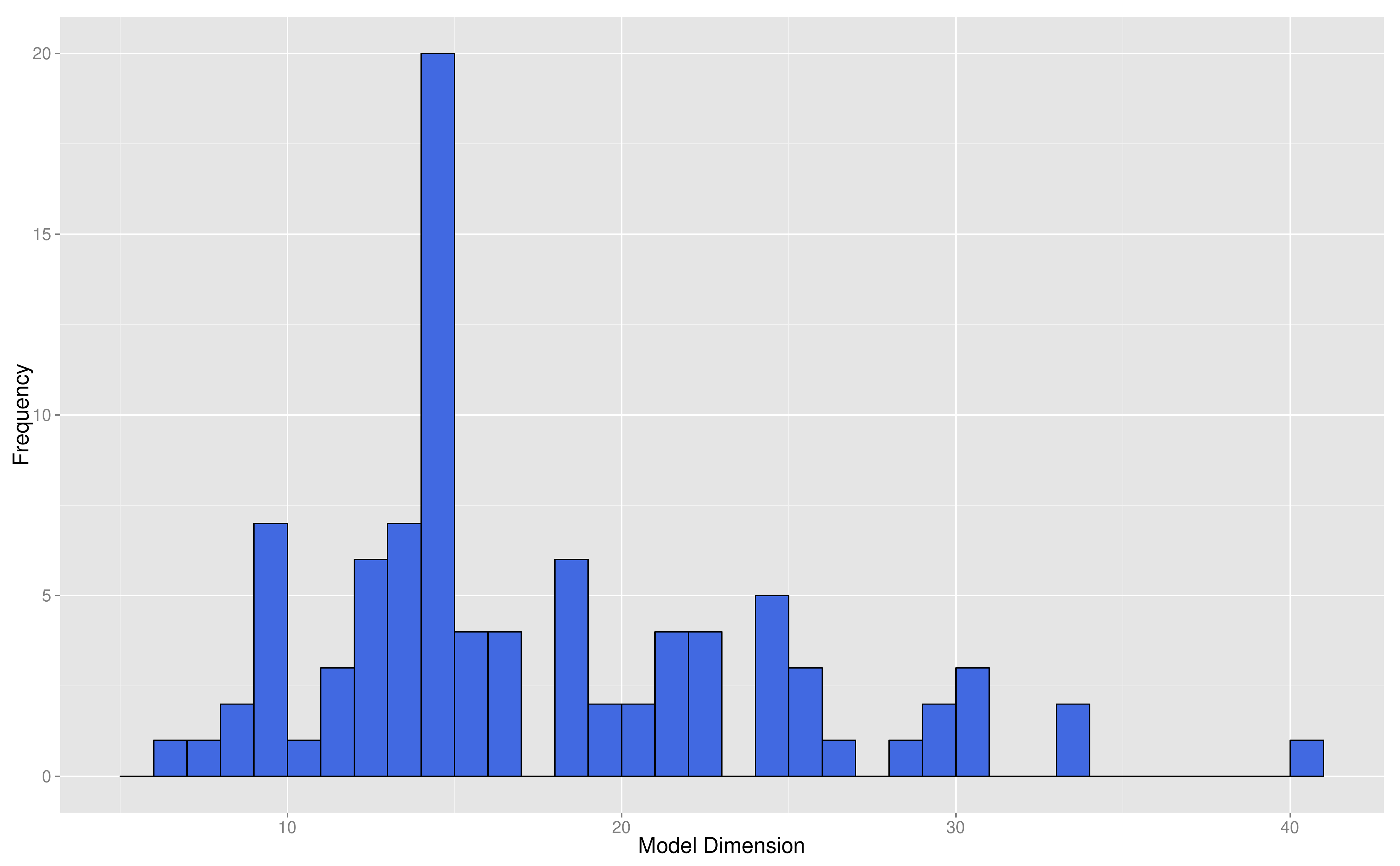}
  \captionof{figure}{Distribution of model dimensionality for all $l =
    92$ signals in the BC under our constrained optimization routine.}
  \label{fig:dicHist}
\end{center}

The histogram in figure~\ref{fig:dicHist} shows the distribution of
$d$ for all $l = 92$ signals in the BC.  It is highly skewed to the
right, with a median (and mode) of 14 PCs and mean of 17 PCs.  We
choose $d = 14$ based on the median of this distribution, and use this
number of explanatory variables in both Bayesian PCR models.  We
choose this as the model that minimizes the risks of both
over-fitting and under-fitting.

\subsection{Inferring the ratio of rotational kinetic energy to
  gravitational energy of the inner core at bounce, $\beta_{ic,b}$}

We injected each of the $l = 92$ BC and $m = 47$ TC signals in
Advanced LIGO noise (SNR $\rho = 20$) and fitted the two Bayesian PCR
models with $d = 14$ PCs.  We then regressed the known value of
$\beta_{ic,b}$ on the posterior means of the BC signals from these
models and sampled from the posterior predictive distribution of the
TC signals.
Figures~\ref{fig:betaKnownT}--\ref{fig:betaUnknownT_Other} show these
predictions of $\beta_{ic,b}$.  The true value from the TC (red
triangle) is compared with the predicted value (blue circle) and
uncertainty is measured using 90\% credible intervals (black lines).
Figures~\ref{fig:betaKnownT}~and~\ref{fig:betaKnownT_Other} assume a
\textit{known} signal arrival time.  $T$ is \textit{unknown} for
figures~\ref{fig:betaUnknownT}~and~\ref{fig:betaUnknownT_Other}.  The
change in background gradient for
figures~\ref{fig:betaKnownT}~and~\ref{fig:betaUnknownT} represents the
varying precollapse differential rotation model $A$ for signals with
LS EOS and standard $Y_e(\rho)$ parametrization.  For
figures~\ref{fig:betaKnownT_Other}~and~\ref{fig:betaUnknownT_Other},
the background shade represents GW signals (from a precollapse
differential rotation model $A1$) with a Shen EOS, or
increase/decrease in $Y_e(\rho)$ of $\sim 5\%$.  $\beta_{ic,b}$ is
scaled up by a factor of 100 in these plots.


\begin{center}
  \includegraphics[width=1\linewidth]{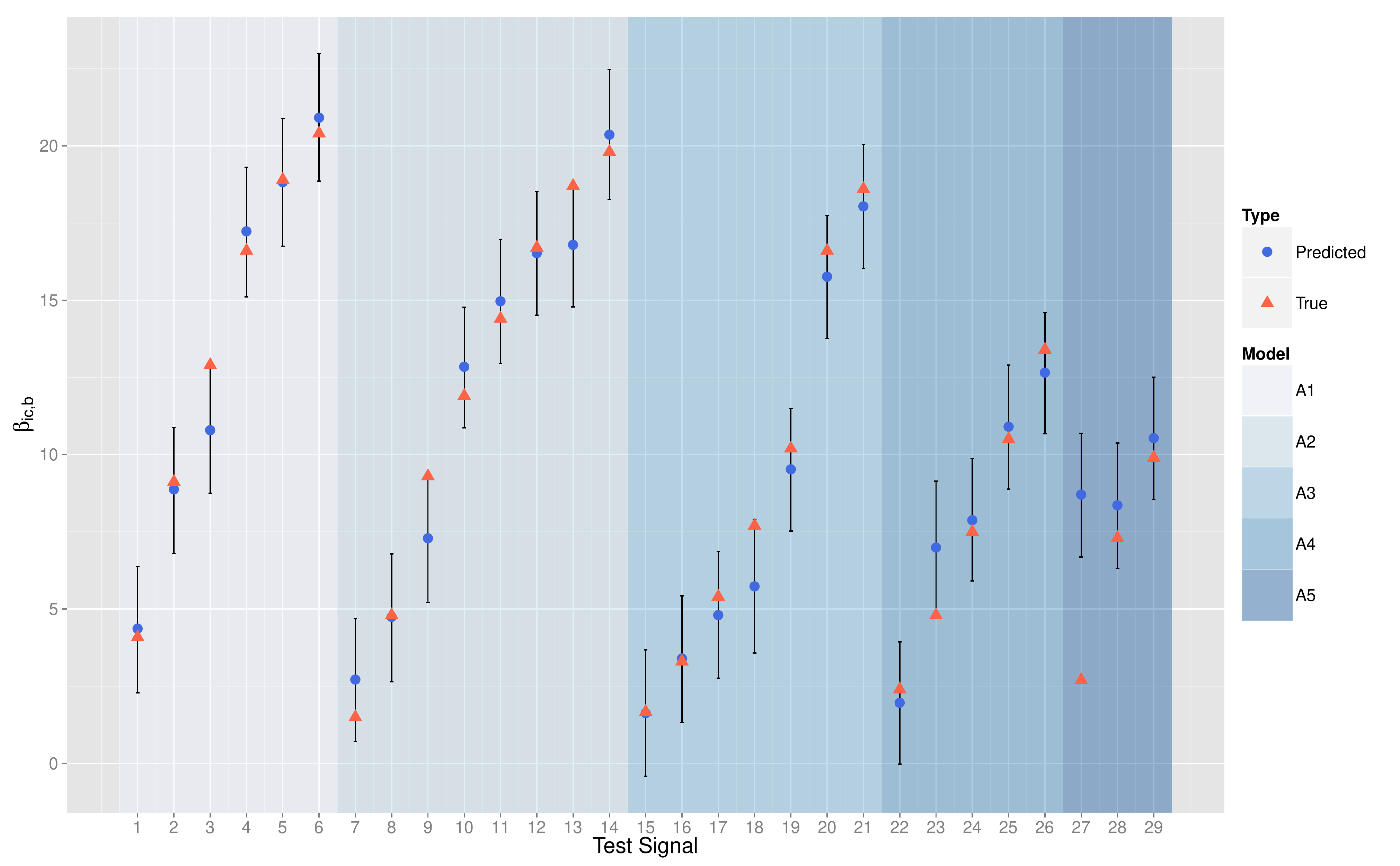}
  \captionof{figure}{90\% credible intervals of $\beta_{ic,b}$ for the
    29 test signals with the LS EOS and standard $Y_e(\rho)$
    parametrization.  $T$ is \textit{known}.}
  \label{fig:betaKnownT}
\end{center}

\begin{center}
  \includegraphics[width=1\linewidth]{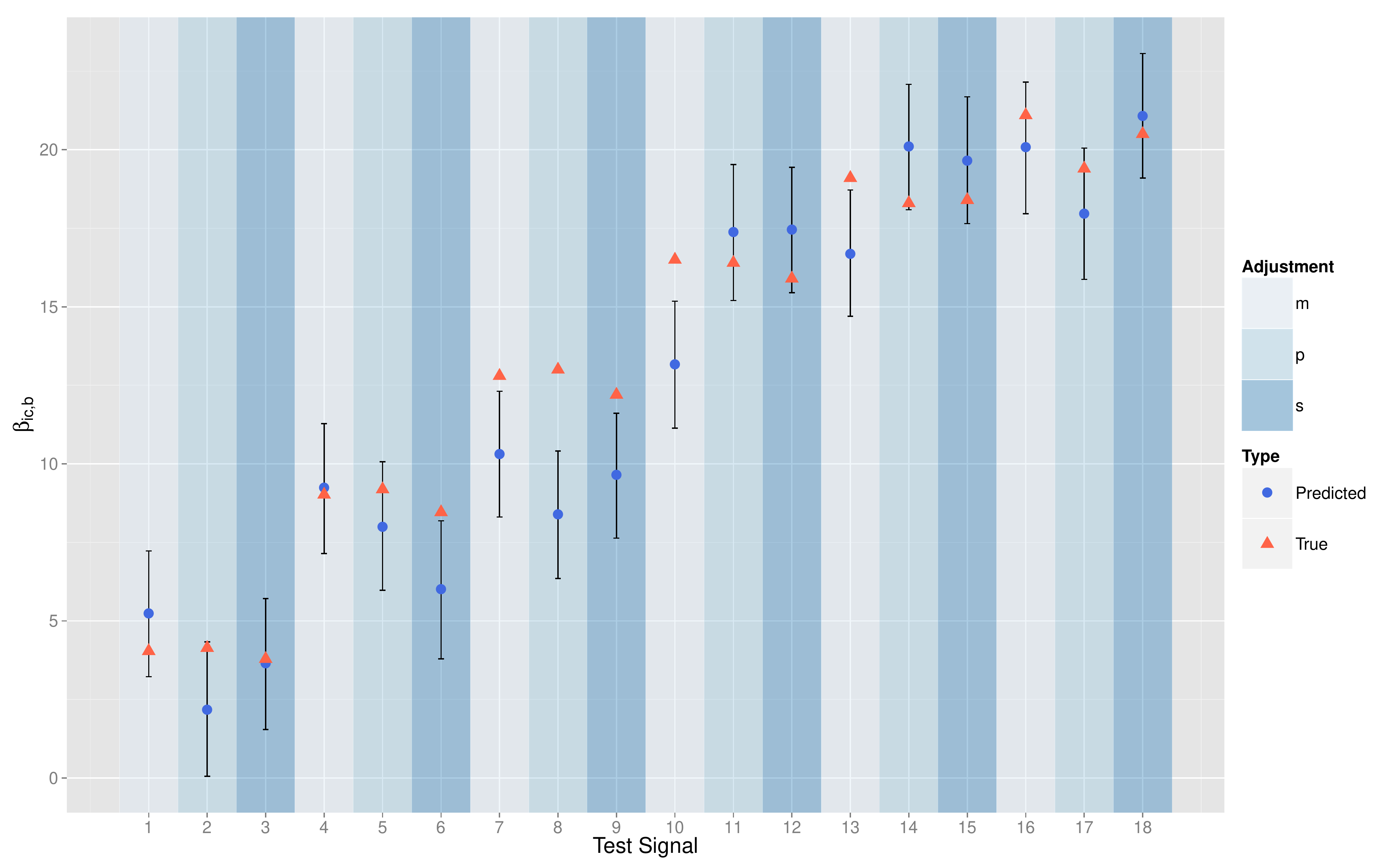}
  \captionof{figure}{90\% credible intervals of $\beta_{ic,b}$ for the
    18 test signals with varying EOS and $Y_e(\rho)$
    parametrization. Note that $m$ refers to an increase in
    $Y_e(\rho)$ of 5\%, $p$ refers to a decrease in $Y_e(\rho)$ of
    5\%, and $s$ refers to the Shen EOS.  $T$ is \textit{known}.}
  \label{fig:betaKnownT_Other}
\end{center}

\begin{center}
  \includegraphics[width=1\linewidth]{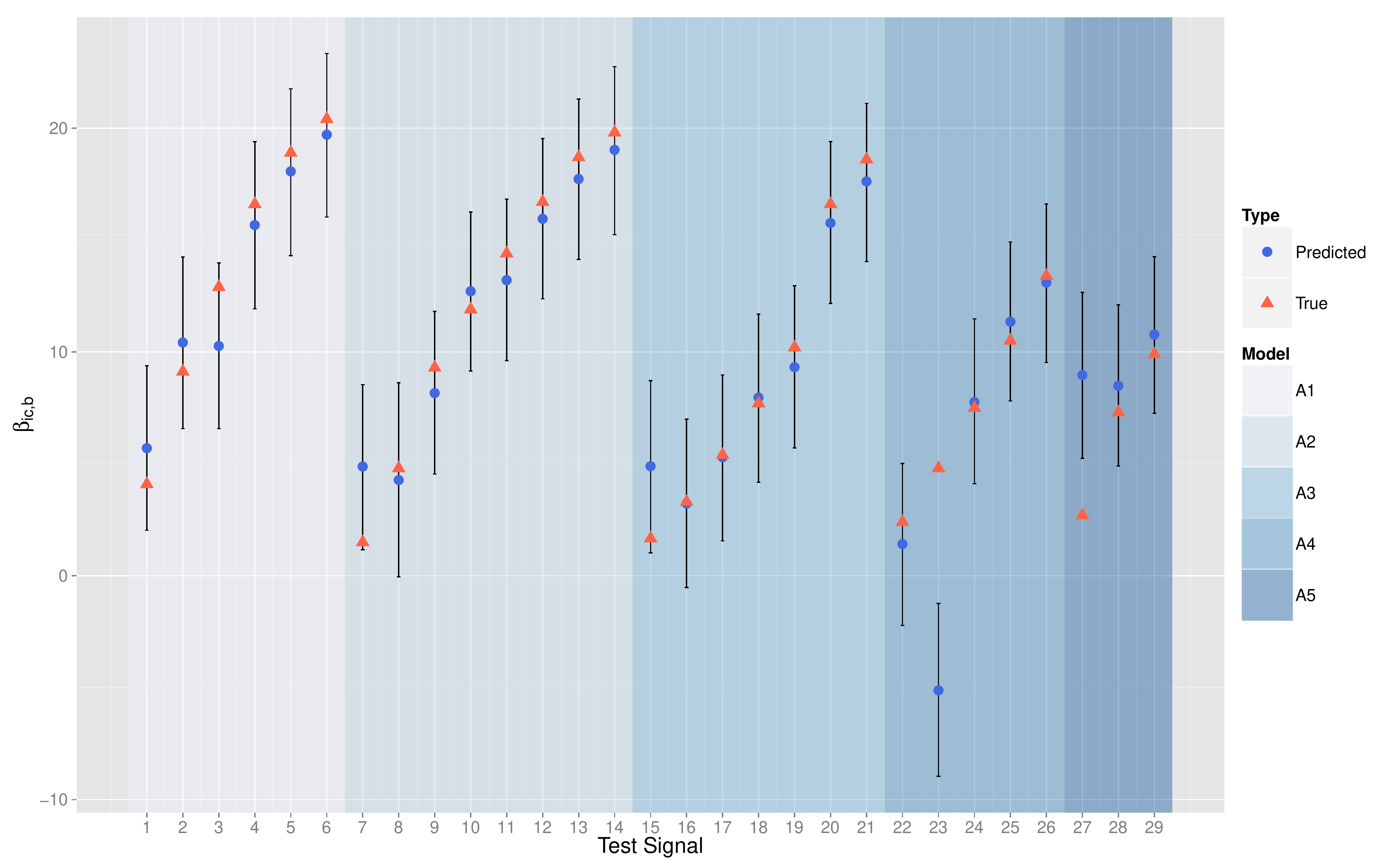}
  \captionof{figure}{90\% credible intervals of $\beta_{ic,b}$ for the
    29 test signals with the LS EOS and standard $Y_e(\rho)$
    parametrization.  $T$ is \textit{unknown}.}
  \label{fig:betaUnknownT}
\end{center}

\begin{center}
  \includegraphics[width=1\linewidth]{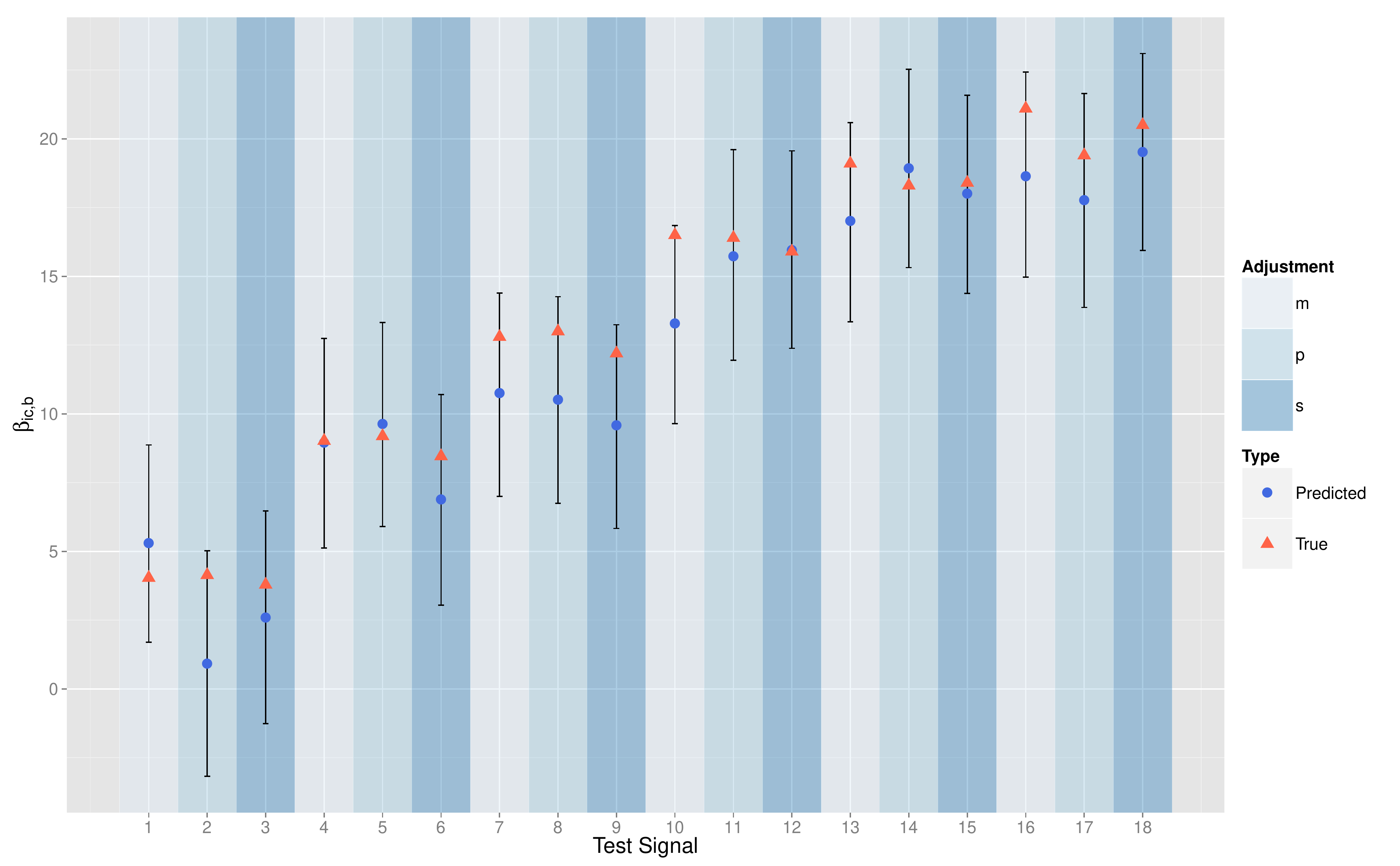}
  \captionof{figure}{90\% credible intervals of $\beta_{ic,b}$ for the
    18 test signals with varying EOS and $Y_e(\rho)$
    parametrization. Note that $m$ refers to an increase in
    $Y_e(\rho)$ of 5\%, $p$ refers to a decrease in $Y_e(\rho)$ of
    5\%, and $s$ refers to the Shen EOS.  $T$ is \textit{unknown}.}
  \label{fig:betaUnknownT_Other}
\end{center}

We yield accurate predictions of $\beta_{ic,b}$ for most of the test
signals in figure~\ref{fig:betaKnownT}.  Signal 27 ($A5O3.25$ from the
catalogue) is an outlier and comes from a slowly rotating core with
uniform rotation.  It is likely an outlier due to the strong
stochastic components in the GW signal from prompt postbounce
convection \cite{abdikamalov:2013}.  The true values of $\beta_{ic,b}$
are on the boundary of the 90\% credible intervals for signals 3
($A1O10.25$), 9 ($A2O6.25$), 19 ($A3O5.25$), and 23 ($A4O3.25$), but
there is no distinguishable pattern between these signals.  The
credible intervals are relatively small, at approximately four units
(times $10^{-2}$) long.  This means that it is particularly easy to
distinguish $\beta_{ic,b}$ between GW signals.

The length of credible interval widens by a factor of $\sim 1.5$ when
changing from known to unknown signal arrival time.  Incorporating an
unknown time shift increases the uncertainty of the PC coefficients
since the MCMC algorithm draws $\boldsymbol\alpha | T$.  That is,
conditioning on an uncertain $T$ creates additional uncertainty for
$\boldsymbol\alpha$.  However, predictions are still accurate in most
cases. We see in figure~\ref{fig:betaUnknownT}, that signal 27
($A5O3.25$) is an outlier again.  Signal 23 ($A4O3.25$) is another
outlier with credible interval on the negative side of the number
line.  This is an absurd and physically impossible range for a
strictly positive variable, and is a consequence of the fact that the
priors can only constrain the linear model parameters
$(\boldsymbol\gamma, \sigma^2)$.  More specifically, we could not put
priors on the response variable of physical parameters
$\boldsymbol\xi$ to constrain the predicted physical parameters
$\check{\boldsymbol\xi}$.  A similarity that this signal has with the
other outlier is that it comes from a slowly rotating core with weak
differential rotation.

Our methods work reasonably well when varying the EOS and
deleptonization parametrization, although we underestimate some
signals with moderate rotation in figure~\ref{fig:betaKnownT_Other}.
Three of these signals come from an increase of $Y_e(\rho)$
parametrization, one from a decrease of $Y_e(\rho)$ parametrization,
and two from the Shen EOS.  When incorporating an unknown time shift
in figure~\ref{fig:betaUnknownT_Other}, the uncertainty of $T$
increases and covers the true parameters.  The increase in the width
of credible interval makes it more difficult to distinguish
$\beta_{ic,b}$ between signals.  

We can conclude that the methods employed in this study are moderately
sensitive to uncertainties in $Y_e(\rho)$ and EOS.  It was found that
a GW signal has relatively weak dependence on the nuclear EOS by
\cite{dimmelmeier:2008}.  We showed in an unpublished study
\cite{edwards:2013} that we could correctly identify between the LS
and Shen EOS for 50\% of the signals in the Dimmelmeier \etal
\cite{dimmelmeier:2008} waveform catalogue using model comparison
techniques. Note that 21\% were incorrectly identified and 29\%
unidentified.  It could therefore be useful to incorporate EOS as an
additional unknown that we wish to infer in future statistical
analyses.

The results presented assume a SNR of $\rho = 20$.  To test
robustness, we trialled the analysis on SNRs of $\rho = 50$ and $\rho
= 100$, which are more realistic levels for detecting CCSN events in
the Milky Way.  Our predictions and credible intervals of
$\beta_{ic,b}$ were the same, regardless of the SNR.  This can be
attributed to using only the posterior means of the PC coefficients in
constructing design matrix $\mathbf{A}$, and not the full spread of
the posterior distributions.  This therefore removes uncertainty due
to LIGO noise and signal reconstruction when predicting $\beta_{ic,b}$
from the posterior predictive distribution.

\subsection{Classifying the precollapse differential rotation, $A$}

Precollapse differential rotation is treated as a categorical variable
with five different levels in this analysis.  We define the set of
classes $C = \{A1, A2, A3, A4, A5\}$ and apply the NBC and $k$-NN
supervised learning algorithms to extract precollapse differential
rotation from each of the signals in the TC.  The model features
$\mathbf{u}$ are the posterior means of the PC coefficients from the
Bayesian PCR models ($\bar{\boldsymbol\alpha}$ for the training set
and $\bar{\check{\boldsymbol\alpha}}$ for the test set). The goal of
this analysis is to let both algorithms learn from the training set to
discriminate GW signals in the test set.

\begin{table}[!h]
  \caption{\label{tab:diffRot} Percentage of signals in the TC with correctly identified precollapse differential rotation $A$ using NBC and $k$-NN.}
    \begin{indented}
    \item[] \begin{tabular}{lcccc}
    \br
    &\multicolumn{2}{c}{Known $T$ (\%)}&\multicolumn{2}{c}{Unknown $T$ (\%)}\\
    \cline{2-5}\noalign{\smallskip}
    Differential Rotation, $A$&NBC&$k$-NN&NBC&$k$-NN\\
    \mr
    $A1$&&&&\\
    -- Standard&83&83&83&83\\
    -- $\uparrow Y_e(\rho)$&67&50&67&50\\
    -- $\downarrow Y_e(\rho)$&67&83&83&100\\
    -- Shen EOS&33&17&0&17\\
    $A2$&50&75&50&50\\
    $A3$&43&57&29&57\\
    $A4$&0&80&20&80\\
    $A5$&33&33&0&33\\
    \br
    \end{tabular}
  \end{indented}
\end{table}

Table~\ref{tab:diffRot} shows the percentage of signals in the TC that
have a correctly identified level of $A$ using NBC and $k$-NN.  We
compare how the methods work when using $\bar{\boldsymbol\alpha}$ and
$\bar{\check{\boldsymbol\alpha}}$ from data with known and unknown
signal arrival times.

The results between models with known and unknown signal arrival times
are quite similar. The standard GWs from class $A1$ are discriminated
well by both algorithms.  The decrease (and to some degree, the
increase) in $Y_e(\rho)$ parametrization did not affect the
algorithms' abilities to discriminate.  Both algorithms performed
particularly poorly for the Shen EOS test signals, which illustrates
that $A$ is sensitive to the EOS.  This is in line with the findings
from \cite{abdikamalov:2013}.

The $k$-NN generally performs better than the NBC for GW signals with
weak to moderate differential rotation ($A3, A4, A5$).  This could be
attributed to our choice in prior classes for the NB method.  Since
models with stronger differential rotation are more populated in the
BC, they have a higher prior probability than those with weaker
differential rotation.

\section{Discussion}

We have presented a Bayesian framework for inferring the physical
parameters of CCSN from GW data.  We have shown
that with a SNR of $\rho = 20$ and optimal orientation of detector to
source, we can extract $\beta_{ic,b}$ with reasonable levels of
uncertainty for the majority of injected test signals.  Both of the
Bayesian PCR models presented in this paper worked well.  The level
of uncertainty increased when incorporating an unknown signal arrival
time into the model, but this is no surprise as PC coefficients are
conditioned on the signal arrival time for that model.  Further, we
found that our methods were moderately sensitive to varying EOS and
$Y_e(\rho)$ parametrizations, and predictions are generally good.

The chosen measure of uncertainty in this analysis was the 90\%
credible interval.  A great benefit of the Bayesian framework is the
probabilistic interpretation of credible intervals, enabling one to
make statements such as, ``with probability 0.9, $\beta_{ic,b}$ is
between $2.5 \times 10^{-2}$ and $6.5 \times 10^{-2}$.''

A true strength of the methods presented in this paper is their
generality.  We initially applied these techniques to the Dimmelmeier
\etal catalogue \cite{dimmelmeier:2008} as a proof of concept and then
easily transferred to the Abdikamalov \etal catalogue
\cite{abdikamalov:2013}. In this study we sampled $\beta_{ic,b}$ from
its posterior predictive distribution.  This method could however be
conducted on any continuous variable of physical interest.  Although
not presented here, predictions of the initial central angular
velocity $\Omega_c$ were comparable to what we found with
$\beta_{ic,b}$.

Choosing to only use the posterior means of the PC coefficients
$\bar{\boldsymbol\alpha}$ in the construction of the design matrix
$\mathbf{A}$ removed some of the variability due to LIGO noise and
signal reconstruction.  The uncertainty from the Bayesian PCR
modelling step therefore does not flow onto the posterior predictive
sampling step.  A more realistic case would be to incorporate this
uncertainty through an errors-in-variables model, which is commonly
used when there are measurement errors in the explanatory variables of
a regression model.  We plan to explore this in a future study.
However, a benefit of our method was that predictions were essentially
independent of SNR (at least for $\rho \geq 20$).

An important task in Bayesian analysis is specifying the prior PDF to
describe our beliefs about model parameters before observing the data.
We wanted to avoid using information from the waveform catalogue as
both data and prior knowledge. It is in this light that we believe the
waveform catalogue should be used only as data, and assume complete
prior ignorance on all of the model parameters.

We applied the NBC and $k$-NN algorithm to extract precollapse
differential rotation.  We found that results were comparable between
known and unknown signal arrival times.  The $k$-NN algorithm
generally performed better than the NBC under the assumptions made.
In future work, we plan to investigate how the choice of prior for the
NBC affects classification, as well as exploring different metrics
such as the Mahalanobis distance (which takes correlations of the data
into account) for the $k$-NN.  We are also investigating an
alternative classification routine, Bayesian ordinal probit
regression.

We introduced a constrained optimization approach to model selection
that allowed us to select an appropriate number of PCs for the
Bayesian PCR models.  To our knowledge, this is the first attempt at
doing so.  Techniques such as reversible jump MCMC (RJMCMC)
\cite{green:1995} have been utilized in GW data analysis contexts
\cite{umstatter:2005}.  RJMCMC could prove to be a useful and more
sophisticated approach than the method presented in the current study.
Although our method required a lot of parallel computing, we found it
to be a novel solution to the model selection problem.

Our analysis assumed optimal orientation of a GW source to a single
interferometer.  As presented in \cite{roever:2007a, roever:2007b} for
compact binary inspiral signals, we plan to extend the methods
presented in this study to a network of detectors.  This is an important
generalization as one can triangulate the position of a GW source
using coherent data from multiple detectors.  The ability to locate a
GW source would allow astronomers to compare and verify whether there
was a true astrophysical event or a glitch with electromagnetic
observations.

\ack

We thank Ik Siong Heng for a thorough reading of the script, Ernazar
Abdikamalov for providing us with the waveform catalogue and
supplementary materials, Christian D. Ott for helpful discussions, and
the New Zealand eScience Infrastructure (NeSI) for their high
performance computing facilities and support. NC's work is supported
by NSF grant PHY-1204371.  This paper has been given LIGO Document
Number P1400034.

\section*{References}

\bibliographystyle{unsrt}


\end{document}